\documentclass{elsart}

\usepackage{amssymb}

\newcommand{\be}{\begin{eqnarray}}
\newcommand{\ee}{\end{eqnarray}}

\newcommand{\Tr}{\,{\rm Tr}\,}

\begin{document}

\begin{frontmatter}

\title{Generalized thermostatistics based on
deformed exponential and logarithmic functions}

\author{Jan Naudts}
\address{
  Departement Natuurkunde, Universiteit Antwerpen,\\
  Universiteitsplein 1, 2610 Antwerpen, Belgium\\
  E-mail {\tt Jan.Naudts@ua.ac.be}
}



\begin{abstract}
The equipartition theorem states that inverse temperature equals
the log-derivative of the density of states.
This relation can be generalized by introducing
a proportionality factor involving an increasing positive function $\phi(x)$.
It is shown that this assumption leads to an
equilibrium distribution of the Boltzmann-Gibbs form with
the exponential function replaced by a deformed exponential function.
In this way one obtains a formalism of generalized thermostatistics
introduced previously by the author. It is shown
that Tsallis' thermostatistics, with a slight modification, is
the most obvious example of this formalism
and corresponds with the choice $\phi(x)=x^q$.
\end{abstract}

\end{frontmatter}

\section{What is thermostatistics?}

See \cite {CHB85} for a conceptual foundation of thermostatistics.
The presentation here emphasizes the role of the density of states.

A model of thermostatistics is described by a density of states
$\rho(E)$ and a probability distribution $p(E)$, both functions
of energy $E$. For a system in thermal equilibrium at temperature $T$,
the probability distribution is
given by the Boltzmann-Gibbs expression
\be
p(E)&=&\frac{1}{Z(T)}e^{-E/T}\qquad\hbox{with}
\label {bg}\\
Z(T)&=&\int{\rm d}E\,\rho(E)e^{-E/T}.
\label {ps}
\ee
(Boltzmann's constant $k_B$ is set equal to 1).
Thermal averages are defined by
\be
\langle f\rangle=\int{\rm d}E\,\rho(E)p(E) f(E).
\label {av}
\ee
Conserved quantities, other than energy, might be important.
It is tradition to give a simplified treatment involving only energy.
To keep notations simple the temperature dependence of $p(E)$
is not made explicit: $p(E)$ stands for $p(E,T)$.

A microscopic model of thermostatistics is specified by an energy
functional $H(\gamma)$ over phase space $\Gamma$, which is the set of all possible
microstates of the system. Using $\rho(E){\rm d}E={\rm d}\gamma$
the integration in expressions (\ref {ps}, \ref {av})
can be replaced by an integration over phase space
\be
\langle f\rangle&=&\int_\Gamma{\rm d}\gamma\,p(\gamma)f(\gamma)
\qquad\hbox{with}\\
p(\gamma)&=&\frac{1}{Z(T)}\exp(-H(\gamma)/T)\qquad\hbox{and}\\
Z(T)&=&\int_\Gamma{\rm d}\gamma\,\exp(-H(\gamma)/T).
\ee
In the quantum case the integration is replaced by a trace over operators.
The formulas are
\be
\langle f\rangle&=&\frac{1}{Z(T)}\Tr \exp(-H/T) f
\qquad\hbox{with}\\
Z(T)&=&\Tr\exp(-H/T).
\ee

Paraphrasing the words of C. Tsallis during this conference,
one can state that the Boltzmann-Gibbs distribution is not just an exponential distribution,
but that it is important that it is a sum over all states of phase space,
and that the exponential contains energy divided by temperature.
In particular, if a microscopic model reproduces the experimentally observed probabilities
$p(\gamma)$ at one given temperature
then one can predict their value at all other temperatures.
This predictive power is the main asset of thermostatistics.

\section{Why Boltzmann-Gibbs?}

In relevant examples of thermostatistics the density of states $\rho(E)$
increases as a power law $\rho(E)\sim E^{\alpha N}$ with $N$ the number of
particles and with $\alpha>0$.
This increase in the density of states is essential to understand
the paradox that according to the Boltzmann-Gibbs distribution
the ground state is always the most probable state. Still, one never
observes that the molecules of the air in a class room lie all on the floor.
Moreover, energy fluctuations of a gas in equilibrium, away from any
phase transition temperature, are negligible.
The solution to the paradox is known as the entropy-energy balance.
The increase of density of states $\rho(E)$ compensates
the exponential decrease of probability density $p(E)$.
The maximum of the product $\rho(E)p(E)$ is reached at some macroscopic
energy far above the ground state energy. Indeed, one can write
\be
\rho(E)p(E)=\frac{1}{Z(T)}\exp\big(\log\rho(E)-E/T\big).
\ee
The argument of the exponential function is maximal if $E$ satisfies
\be
\frac{1}{\rho(E)}\rho'(E)=\frac{1}{T}
\label {eqpthm}
\ee
($\rho'(E)$ is the derivative of $\rho(E)$ w.r.t.~$E$).
If $\rho(E)\sim E^{\alpha N}$ then $E\simeq\alpha N T$ follows,
which is the equipartition theorem.

As a consequence of the equipartition theorem it is not
easy to verify the Boltzmann-Gibbs distribution experimentally.
Indeed, the energy of the system under study is always equal
to the value predicted by (\ref {eqpthm}), with negligible
fluctuations. This indicates that the actual form of the
probability distribution $p(E)$ is not firmly established and, in fact,
is not very essential. Alternative expressions for $p(E)$
are acceptable provided they satisfy the equipartition theorem
and reproduce thermodynamics. An indication
of the need for a generalization of Boltzmann-Gibbs is the
ubiquitous use of temperature-dependent Hamiltonians in
applied statistical physics. As stressed in the previous section,
(\ref {bg}) predicts the probability density $p(E)$ at all
temperatures. In many cases this prediction is not very
accurate, probably because of an incomplete knowledge
of the density of states $\rho(E)$. However, instead
of making $\rho(E)$ temperature-dependent, which is not
supported by theory, one can as well try to replace the
Boltzmann-Gibbs distribution by another expression
more appropriate for the problem at hand.

\section{The basic postulate}

The present generalization of thermostatistics starts with
generalizing the equipartition result (\ref {eqpthm}).
Let us postulate the existence of an increasing positive function $\phi(x)$,
defined for $x\ge 0$, such that
\be
\frac{1}{T}=\frac{-p'(E)}{\phi\big(p(E)\big)}
\label {phidef}
\ee
holds for all energies $E$ and temperatures $T$.
Then the equation for the maximum of $\rho(E)p(E)$ becomes
\be
0&=&\frac{{\rm d}\,}{{\rm d}E}\bigg(\rho(E)p(E)\bigg)\cr
&=&\rho'(E)p(E)
-\frac{1}{T}\rho(E)\phi\big(p(E)\big).
\ee
This can be written as
\be
\frac{\rho'(E)}{\rho(E)}
=\frac{1}{T}\frac{\phi\big(p(E)\big)}{p(E)}.
\ee
The latter expression generalizes the equipartition theorem (\ref {eqpthm}).
The Boltz\-mann-Gibbs case is recovered with $\phi(x)=x$.

The postulate (\ref {phidef}) fixes the form of the probability
distribution $p(E)$. To see this,
introduce a function, denoted $\ln_\phi(x)$, by
\be
\ln_\phi(x)=\int_1^x{\rm d}y\,\frac{1}{\phi(y)}.
\ee
There are good reasons for calling this function
a deformed logarithm. If $\phi(x)=x$ then it
coincides with the natural logarithm $\log(x)$.
Because $\phi(x)$ is positive for all positive $x$ one has that
$\ln_\phi(x)$ is negative for $0<x<1$ and positive for $x>1$.
With some further technical conditions this function
becomes a deformed logarithm in the sense of \cite {NJ02}.

The inverse of the function $\ln_\phi(x)$ is denoted $\exp_\phi(x)$.
From the identity $1=\exp'_\phi\big(\ln_\phi(x)\big)\ln'_\phi(x)$
follows
\be
\phi(x)=\exp'_\phi\big(\ln_\phi(x)\big).
\label {phiexpr}
\ee
Hence (\ref {phidef}) can be written as
\be
p'(E)=-\frac{1}{T}\exp'_\phi\big[\ln_\phi\big(p(E)\big)\big].
\ee
This expression can be integrated. The result is
\be
p(E)=\exp_\phi\big(G_\phi(T)-E/T\big).
\label {genpdf}
\ee
The function $G_\phi(T)$ is the integration constant. It must be
chosen in such a way that the normalization condition
\be
1=\int{\rm d}E\, \rho(E)p(E)
\label {norm}
\ee
is satisfied. Positivity $p(E)\ge 0$ is automatic because the
range of the function $\exp_\phi(x)$ is the domain of
$\ln_\phi(x)$, with possibly $0$ and $+\infty$ added.
Expression (\ref {genpdf}) resembles the Boltzmann-Gibbs distribution
(\ref {bg}). An important difference is that the normalization
constant appears inside the function $\exp_\phi(x)$.
In case $\phi(x)=x$ then one has $G_\phi(T)=-\log\big(Z(T)\big)$.

Starting from (\ref {genpdf}) a generalized thermostatistics can be
developed --- see \cite {NJ03}. Most of the results that
follow below are reformulations of the results found in the first part of \cite {NJ03}.
The variational principle obeyed by (\ref {genpdf}) is
not discussed below.

\section{Escort probabilities}

In general it is difficult to calculate the integration constant $G_\phi(T)$.
A useful expression for its temperature derivative can be obtained in
terms of {\sl escort probabilities}. They originate from \cite {BS93} and
have been introduced in Tsallis' thermostatistics in \cite {TMP98}.
The generalized definition is
\be
P(E)=\frac{1}{Z(T)}\phi\big(p(E)\big)
\label {Pdef}
\ee
with normalization factor
\be
Z(T)=\int{\rm d}E\,\rho(E)\phi\big(p(E)\big).
\ee
Expectation values w.r.t.~$P(E)$ are denoted
\be
\langle f\rangle_*=\int{\rm d}E\,\rho(E)P(E) f(E).
\label {escav}
\ee
Note that $P(E)$ coincides with $p(E)$
in the Boltzmann-Gibbs case $\phi(x)=x$ for all $x$.

Now calculate, using (\ref {phiexpr}) and (\ref {Pdef}),
\be
\frac{{\rm d}\,}{{\rm d}T}p(E)
&=&\exp'_\phi\big(G_\phi(T)-E/T\big)\left(\frac{{\rm d}\,}{{\rm d}T}G_\phi(T)+\frac{E}{T^2}\right)\cr
&=&Z(T)P(E)\left(\frac{{\rm d}\,}{{\rm d}T}G_\phi(T)+\frac{E}{T^2}\right).
\label {pder}
\ee
From (\ref {norm}) and (\ref {pder}) follows
\be
0&=&\int{\rm d}E\, \rho(E)\frac{{\rm d}\,}{{\rm d}T}p(E)\cr
&=&Z(T)\frac{{\rm d}\,}{{\rm d}T}G_\phi(T)+\frac{1}{T^2}Z(T)\langle E\rangle_*.
\ee
Hence one has
\be
\frac{{\rm d}\,}{{\rm d}T}G_\phi(T)=-\frac{1}{T^2}\langle E\rangle_*.
\label {dgdt}
\ee
For further use note that (\ref {pder}) and (\ref {dgdt}) together give
\be
\frac{{\rm d}\,}{{\rm d}T}p(E)
&=&\frac{1}{T^2}Z(T)P(E)\left(E-\langle E\rangle_*\right).
\label {pder2}
\ee

\section{Thermodynamic relations}

One goal of thermostatistics is to give a microscopic derivation
of the laws of thermodynamics. This raises immediately
the question in how far generalized thermostatistics is still
compatible with thermodynamics.

Let us start with establishing thermal stability.
Internal energy $U(T)$ is defined by $U(T)=\langle E\rangle$,
with $p(E)$ given by (\ref {genpdf}).
Using (\ref {pder2}) one obtains
\be
\frac{{\rm d}\,}{{\rm d}T}U(T)
&=&\int{\rm d}E\,\rho(E) E \frac{{\rm d}\,}{{\rm d}T}p(E)\cr
&=&\int{\rm d}E\,\rho(E) E \frac{1}{T^2}Z(T)P(E)\left(E-\langle E\rangle_*\right)\cr
&=&\frac{1}{T^2}Z(T)\left(\langle E^2\rangle_*-\langle E\rangle_*^2\right)\cr
&\ge&0.
\ee
Hence, average energy is an increasing function of temperature. However,
thermal stability requires more.
Define $\phi$-entropy (relative to $\rho(E){\rm d}E$) by
\be
S_\phi(p)
&=&\int{\rm d}E\,\rho(E)\left[\big(1-p(E)\big)F_\phi(0)-F_\phi\big(p(E)\big)\right]
\label {sdef}
\ee
with
\be
F_\phi(x)=\int_1^{x}{\rm d}y\,\ln_\phi(y).
\ee
Let us postulate that thermodynamic entropy $S(T)$ equals the value of
the above entropy functional $S_\phi(p)$ with $p$ given by (\ref {genpdf}).
Then one finds
\be
\frac{{\rm d}}{{\rm d}T}S(T)
&=&\int{\rm d}E\,\rho(E)\big(-\ln_\phi(p(E))-F_\phi(0)\big)\frac{{\rm d}\,}{{\rm d}T}p(E)\cr
&=&\int{\rm d}E\,\rho(E)\left(-G_\phi(T)+\frac{E}{T}-F_\phi(0)\right)\frac{{\rm d}\,}{{\rm d}T}p(E)\cr
&=&\frac{1}{T}\frac{{\rm d}\,}{{\rm d}T}U(T).
\label {dsdt}
\ee
To see the latter use that $p(E)$ is normalized to 1. This shows that temperature $T$
satisfies the thermodynamic relation
\be
\frac{1}{T}=\frac{{\rm d}S}{{\rm d}U}.
\ee
Because energy is an increasing function of temperature one concludes that
entropy $S$ is a concave function of energy $U$. This property is known as
thermal stability.

One can introduce the Helmholtz free energy $F(T)$ by the well-known relation
$F(T)=U(T)-TS(T)$. From (\ref {dsdt}) follows that
\be
\frac{{\rm d}\,}{{\rm d}\beta}\beta F(T)=U(T)
\qquad\hbox{ with }\beta=\frac{1}{T}.
\label{energ}
\ee
Now let us come back to (\ref {dgdt}), which is very similar to (\ref {energ}),
with $F(T)$ replaced by $TG_\phi(T)$ and
with $U(T)=\langle E\rangle$ replaced by $\langle E\rangle_*$.
The comparison shows that $TG_\phi(T)$ equals the free energy associated with
the escort probability distribution $P(E)$,
up to a constant independent of temperature $T$.

\section{Tsallis' thermostatistics}

The most obvious generalization of Boltzmann-Gibbs thermostatistics is obtained by
the choice $\phi(x)=x^q$ with $q>0$. It reproduces Tsallis' thermostatistics
with some minor changes.
The corresponding deformed logarithmic and exponential functions are
\be
\ln_q(x)&=&\int_1^x{\rm d}y\,y^{-q}=\frac{1}{1-q}\left(x^{1-q}-1\right)\cr
\exp_q(x)&=&\left[1+(1-q)x\right]_+^{1/(1-q)}.
\ee
These have been introduced in \cite {TC94}.
The probability distribution (\ref {genpdf}) becomes
\be
p(E)&=&\left[1+(1-q)\big(G_q(T)-E/T\big)\right]_+^{1/(1-q)}
\label {tsallisnew}\\
&=&\frac{1}{z_q(T)}\left[1-(1-q)\beta_q^*(T)E\right]_+^{1/(1-q)}
\label {tsallisorgpdf}
\ee
with
\be
z_q(T)&=&\left(1+(1-q)G_q(T)\right)^{1/(q-1)}
\qquad\hbox{and}\cr
\beta_q^*(T)&=&z_q(T)^{1-q}/T.
\ee
Originally \cite {TC88}, (\ref {tsallisorgpdf}) was introduced with $q$
replaced by $2-q$. The present expression was proposed in \cite {CT91},
be it with a different expression for $\beta_q^*(T)$.

A nice feature of Tsallis' thermostatistics is that the correspondence
between $p(E)$ and the escort $P(E)$ leads to a dual structure,
called the '$q\leftrightarrow 1/q$'-duality \cite {TMP98,NJ01}.
Indeed, from
\be
P(E)=\frac{1}{Z_q(T)}p(E)^q
\qquad\hbox { follows }\quad
p(E)=\frac{1}{Z_{1/q}(T)}P(E)^{1/q}.
\ee
However, there exists also a '$q\leftrightarrow 2-q$'-duality
the origin of which has been clarified in \cite {NJ02}.
Given $\ln_\phi(x)$, a new deformed logarithmic function $\ln_\psi(x)$
is obtained by
\be
\ln_\psi(x)=(x-1)F_\phi(0)-xF_\phi(1/x),
\label {dedlog}
\ee
with $\psi(x)$ given by
\be
\frac{1}{\psi(x)}=F_\phi(0)-F_\phi(1/x)+\frac{1}{x}\ln_\phi\left(\frac{1}{x}\right).
\ee
In case $\phi(x)=x^q$ follows $\psi(x)=(2-q)x^{2-q}$. Hence,
the deformed logarithms $\ln_q(x)$ and $\ln_{2-q}(x)$
can be deduced from each other, up to a constant factor, by the relation (\ref {dedlog}).

Now note that the definition (\ref {sdef}) of entropy $S_\phi(p)$ can be written as
\be
S_\phi(p)=\int{\rm d}E\,\rho(E)p(E)\ln_\psi\big(1/p(E)\big).
\ee
With $\psi(x)=x^q$ this expression is Tsallis' entropy
\be
S_q(p)=\int{\rm d}E\,\rho(E)\frac{1}{1-q}\left(p(E)^q-p(E)\right).
\label {tsalentr}
\ee
However,
as noted above, $\psi(x)\sim x^{2-q}$ is needed. This leads to
an expression for entropy with $q$ replaced by $2-q$.
In the Tsallis literature one has preferred to use always the
same expression (\ref {tsalentr}) for the entropy functional.
Instead, the definition of average energy has been
changed \cite {TMP98} from $\langle E\rangle$ to $\langle E\rangle_*$.
For most physicists the latter change is unacceptable.
During this conference F. Baldovin suggested to interchange
the roles of $p(E)$ and $P(E)$ so that the need for changing
the definition of average energy disappears. This is what happened in the present
paper.

\section{Conclusions}

It is possible to formulate a generalized thermostatistics in which
the equipartition theorem holds in a modified form. Some aspects
of thermodynamics, like thermal stability, are recovered. Other
aspects, like extensivity, have not been considered because
they will not hold in the generalized context. In the presentation,
the role of the density of states $\rho(E)$ has been emphasized.
By doing so the link is made between probability distribution functions
$p(E)$ depending on energy $E$ and the distributions $p(\gamma)$
depending on points $\gamma$ in phase space $\Gamma$.

The generalized formalism is determined by the choice of an increasing positive
function $\phi(x)$, defined for $x\ge 0$. The standard formalism
corresponds with $\phi(x)=x$ for all $x$. Tsallis' thermostatistics,
with some minor changes, corresponds with the obvious choice $\phi(x)=x^q$.
The function $\phi(x)$ can be used to define deformed logarithmic
and exponential functions. The equilibrium probability distribution function
$p(E)$ is then a generalization of the Boltzmann-Gibbs distribution, obtained by
replacing the exponential function $\exp(x)$ by the deformed exponential
function.

One of the advantages of formulating a rather general theory is that
it clarifies some aspects of Tsallis' thermostatistics by putting them
in a broader context. In particular, the '$q\leftrightarrow 1/q$' and
'$q\leftrightarrow 2-q$'-dualities have been discussed. The former one
concerns the role of escort probabilities. The latter concerns
the observation that from each deformed logarithm one can deduce another deformed logarithm.
Clarifying these points makes clear that there is no need to
modify the definition of average energy, as has been done in the recent
Tsallis literature. Rather one should replace Tsallis' entropy $S_q(p)$
by $S_{2-q}(p)$. In this way one obtains a formalism which is satisfactory
both from a physical and from a mathematical point of view, and which
can be generalized to a much broader scope when needed.


\end{document}